\documentclass[aps,prl,twocolumn,amsmath,amssymb,nofootinbib,superscriptaddress]{revtex4}
\usepackage[english]{babel}
\usepackage{latexsym}
\usepackage{graphics}
\usepackage{epsfig}
\usepackage{color}
\usepackage{bm}
\usepackage{amsmath}
\usepackage{amssymb}
\usepackage{subfig}

\begin{document}

\title{Stable $p$-Wave Resonant Two-Dimensional Fermi-Bose Dimers}
\author{B.~Bazak}
\affiliation{The Racah Institute of Physics, The Hebrew University, 9190401, Jerusalem, Israel}

\author{D.~S.~Petrov}
\affiliation{LPTMS, CNRS, Univ. Paris Sud, Universit\'e Paris-Saclay, 91405 Orsay, France}

\date{\today}

\begin{abstract}

We consider two-dimensional weakly bound heterospecies molecules formed in a Fermi-Bose mixture with attractive Fermi-Bose and repulsive Bose-Bose interactions. Bosonic exchanges lead to an intermolecular attraction, which can be controlled and tuned to a $p$-wave resonance. Such attractive fermionic molecules can be realized in quasi-two-dimensional ultracold isotopic mixtures. We show that they are stable with respect to the recombination to deeply bound molecular states and with respect to the formation of higher-order clusters (trimers, tetramers, etc.)

\end{abstract}

\maketitle

One of the most paradigmatic examples of a topological quantum system is the $p_x+ip_y$ superfluid, realized in $^3$He \cite{Volovik1992} and possibly in superconducting Sr$_2$RuO$_4$ \cite{Kallin2012}. In spite of these observations, there is a constant search for more robust and controllable setups with a better access to the interesting topological properties of this nontrivial phase \cite{ReadGreen2000,Volovik2003}, Majorana modes, non-Abelian vortices, and, eventually, to the topologically protected quantum computing \cite{NayakRMP2008}. Ultracold gases make a promising platform for this search because of their purity, controllability, and successful past performance, particularly resulted in the comprehensive characterization of the crossover from the $s$-wave Bardeen-Cooper-Schrieffer (BCS) pairing to the Bose-Einstein condensation of molecules in two-component Fermi gases \cite{BlochRMP2008,GiorginiRMP2008,KetterleVarenna2008,ZwergerVolume2011,LevinsenParish2015}. 

The analogous crossover for spinless fermions with $p$-wave attraction is expected to be fundamentally different and more challenging for both theory and experiment. By using a variational BCS-type ansatz it has been shown that with increasing the attraction the system should cross a transition from the weakly coupled topological $p_x+ip_y$ phase to the topologically trivial strongly coupled phase of dimers (see \cite{GurarieRadzihovsky2007} for review). However, inclusion of three-body and possibly higher-order correlations may significantly modify this picture \cite{Levinsen2007,JonaLasinio2008,NishidaSuperEfimov2013,Volosniev2014,Gao2015,Zhang2017}. From the experimental side, in spite of the availability of $p$-wave Feshbach resonances for fermionic alkalis \cite{Regal2003,Zhang2004,Ticknor2004,Gunter2005,Schunck2005,Gaebler2007,Inada2008,Fuchs2008,Nakasuji2013,Waseem2016,Luciuk2016}, these systems suffer from enhanced three-body losses; the two-body wave function near a $p$-wave resonance is localized at short interparticle distances where the pair is ``preformed'' and can easily recombine to deeper molecular states when the third particle comes nearby \cite{Levinsen2007,JonaLasinio2008,Levinsen2008}. Reaching the $p$-wave superfluidity while keeping inelastic losses under control is one of the most challenging problems in the field \cite{Zhang2008,Han2009,Cooper2009,Levinsen2011,JuliaDiaz2013,Buchler2014,Yamaguchi2015,Fedorov2017}. 

The inelastic decay can be suppressed, if the support of the $p$-wave attraction extends much beyond the recombination region. A remarkable experimental demonstration of this phenomenon, although not for identical fermions, is the strong and recombination-free $p$-wave attraction between $^{40}$K atoms and weakly bound $^{40}$K-$^6$Li molecules, driven by the Li atom exchange \cite{Jag2014}. Here, the range of the atom-molecule potential, comparable to the size of the molecule, is much larger than the interatomic van der Waals range, which determines the recombination region. The idea of separating the two scales is also compatible with numerous theoretical proposals of achieving the $p$-wave pairing by immersing {\it identical} fermionic impurities into a Bose or Fermi gas of another species, the attraction between the impurities being generated by exchanges of phonons or particle-hole excitations in the host gas \cite{KaganChubukov1988,EfremovViverit2002,Matera2003,MurPetit2004,Wang2005,Bulgac2006,Suzuki2008,Yang2008,Nishida2009,Bulgac2009,Dutta2010,
Massignan2010,Matera2011,Kain2012,Lai2013,Liao2013,Midtgaard2016,Wu2016,Caracahnas2017,Kelly2018}.

In this Letter, we consider weakly bound two-dimensional Fermi-Bose molecules and investigate their scattering properties. We show that these composite fermions experience a $p$-wave attraction, which strengthens with the Fermi-Bose mass ratio $m_F/m_B$ and weakens with the ratio of the Bose-Bose (BB) to Fermi-Bose (FB) two-dimensional scattering length $a_{\rm BB}/a_{\rm FB}$ \cite{LL}. A finite BB repulsion is necessary to avoid the formation of Fermi-Bose-Bose (FBB) and higher-order clusters in collisions of two or larger number of molecules. We find that under current experimental conditions the attraction between molecules can be made sufficiently strong and their lifetime sufficiently long for observing the $p$-wave pairing, particularly, in the quasi-two-dimensional isotopic $^{40}$K-$^{39}$K mixture.

The mechanism of the molecule-molecule attraction can be qualitatively understood as follows. Consider two identical fermionic atoms interacting with a boson via a short-range potential that supports a weakly bound FB molecular state. The exchange of the boson leads to an effective attraction, inversely proportional to the boson mass $m_{\rm B}$  \cite{Fonseca,LesHouches,Ngampuertikorn}. On the other hand, the fermionic quantum statistics imposes an effective centrifugal repulsion $\propto l(l+1)/m_{\rm F}R^2$ in three dimensions and $\propto l^2/m_{\rm F}R^2$ in two dimensions, where $l$ is an odd integer and $R$ is the distance between the fermionic atoms. Above a critical mass ratio $(m_{\rm F}/m_{\rm B})_c$, which marks an atom-molecule resonance, the exchange attraction overcomes the centrifugal barrier and there appears a trimer state with angular momentum $l$. In three dimensions the $l=1$ trimer appears above $(m_{\rm F}/m_{\rm B})_c=8.2$ \cite{KartavtsevMalykh2007} and in two dimensions this number decreases to $(m_{\rm F}/m_{\rm B})_c=3.3$ \cite{PricoupenkoPedri2010}, consistent with the factor of 2 reduction of the centrifugal barrier. Below $(m_{\rm F}/m_{\rm B})_c$ the system is characterized by an atom-molecule attraction in the $p$-wave channel, like the one observed in the three-dimensional mixture of $^{40}$K atoms and $^{40}$K-$^6$Li molecules \cite{Jag2014,remark1}.

Now consider two FB molecules (FFBB system), for the moment neglecting the BB interaction. Then, the presence of the second boson approximately doubles the exchange attraction and we expect a $p$-wave molecule-molecule resonance to appear at a roughly twice lower mass ratio than in the FFB case. There are problems with this picture, particularly in the three-dimensional case where the FBB subsystem is Efimovian \cite{Efimov1973,BraatenHammer2006,Helfrich2010} and, therefore, features a ladder of FBB trimer states and an enhanced local three-body FBB correlator. Both these factors make molecules prone to inelastic losses. Although there may be a way to minimize losses \cite{Naidon2018}, we turn our attention to the (non-Efimovian) two-dimensional case. In this case, the above estimates suggest that the molecule-molecule $p$-wave interaction becomes resonant for $1\lesssim m_{\rm F}/m_{\rm B}\lesssim 2$ offering a possibility to study effects of strong $p$-wave interactions in accessible isotopic mixtures.   

In the rest of the Letter we focus on the cases $m_{\rm F}/m_{\rm B}=1$ and $m_{\rm F}/m_{\rm B}=2$. We first investigate the energetic stability of two- and three-molecule collisions with respect to the formation of clusters of the type F$_N$B$_M$ and determine the molecule-molecule scattering properties in the purely two-dimensional case. Then, we apply our results to a realistic quasi-two-dimensional setup and discuss relaxation losses.

\begin{figure*}[hptb]
  \begin{center}
    \includegraphics[width=1.023\columnwidth]{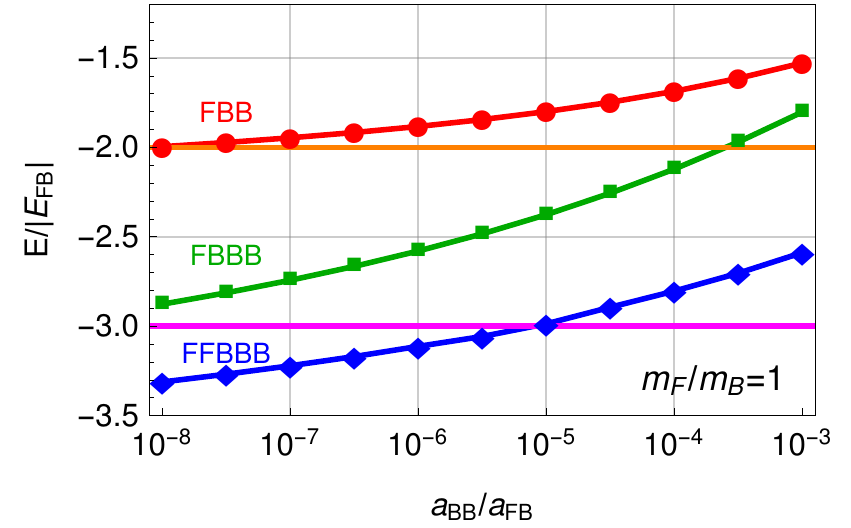}
    \includegraphics[width=0.977\columnwidth]{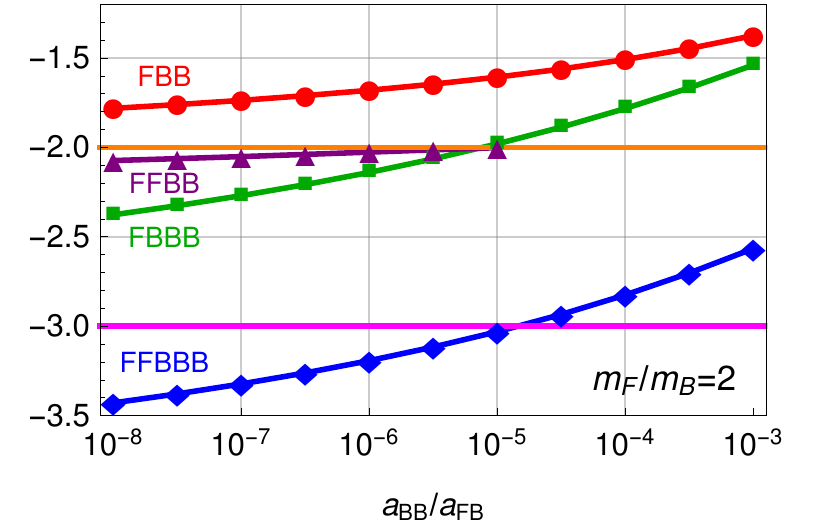}
\caption{Energies of various Fermi-Bose clusters in units of the molecule energy $E_{\rm FB}$ for $m_{\rm F}/m_{\rm B}=1$ (left panel) and $m_{\rm F}/m_{\rm B}=2$ (right panel). The solid curves are linear interpolations of the data. The orange and pink horizontal lines indicate the two- and three-molecule scattering thresholds, respectively. The FFBB tetramer exists in the case $m_{\rm F}/m_{\rm B}=2$. Its crossing with the two-molecule threshold marks the molecule-molecule $p$-wave resonance.}
\vspace{-0.5cm}
\label{Fig:Spectra}
\end{center}
\end{figure*}

We model the two-dimensional FB and BB interactions by, respectively, attractive and repulsive Gaussian potentials $\propto \exp(-r^2/2r_0^2)$. The FB potential is so shallow that the FB scattering length $a_{\rm FB}$ is much larger than the range $r_0$ and there is a weakly bound molecular state with the energy close to the universal zero-range value $E_{\rm FB}=-2e^{-2\gamma}/\mu a_{\rm FB}^2$, where $\gamma=0.5772$ is Euler's constant, $\mu=m_{\rm F}m_{\rm B}/(m_{\rm F}+m_{\rm B})$, and we set $\hbar=1$. The repulsive BB potential supports no bound states. It is quantified by $0<a_{\rm BB}\lesssim r_0$. 

A low-energy collision between two FB molecules can result in the creation of FFB or FBB trimers. Their energies in two dimensions have been calculated for any mass ratio in the absence of the BB interaction \cite{PricoupenkoPedri2010} (see also \cite{Brodsky2006,Bellotti2011}). The FFB trimer does not exist for $m_{\rm F}/m_{\rm B}<3.3$ \cite{PricoupenkoPedri2010}. By contrast, the FBB trimer exists and its energy is always below the molecule-molecule collision threshold $2E_{\rm FB}$. However, we show (see Fig.~\ref{Fig:Spectra}) that already a weak BB repulsion ($a_{\rm BB}/a_{\rm FB}>7.65\times 10^{-9}$ for $m_{\rm F}/m_{\rm B}=1$ and $a_{\rm BB}/a_{\rm FB}>3.2\times 10^{-22}$ for $m_{\rm F}/m_{\rm B}=2$) is sufficient to push the FBB trimer up and thus energetically forbid its formation in low-energy molecule-molecule collisions.

Three-molecule collisions can, in principle, result in the formation of the fermion-rich FFFB and FFFBB and boson-rich FBBB and FFBBB clusters. The FFFB tetramer is not bound \cite{LevinsenParish2013} and we find no FFFBB pentamer for the considered mass ratios. By contrast, the boson-rich clusters exist and we show their ground-state energies in Fig.~\ref{Fig:Spectra} (FB$_M$ and F$_2$B$_M$ clusters have, respectively, zero and unit angular momenta). We observe that as long as the trimer-formation channel FB+FB$\rightarrow$ FBB+F is energetically closed, the reaction FB+FB+FB$\rightarrow$FBBB+F+F is also energetically forbidden. However, the reaction FB+FB+FB$\rightarrow$FFBBB+F requires a stronger BB repulsion to become energetically forbidden. The corresponding critical ratios $a_{\rm BB}/a_{\rm FB}$ for $m_{\rm F}/m_{\rm B}=1$ and $m_{\rm F}/m_{\rm B}=2$ equal 
$8 \times 10^{-6}$ and $1.4 \times 10^{-5}$, respectively. 
Thus, in this two-dimensional model, for a sufficiently strong BB repulsion, two- and three-molecule collisions are elastic. We attribute the relatively small critical values of the BB interaction to the approximate balance between centrifugal and boson-exchange forces for the considered mass ratios. The analysis of inelastic channels in $N$-molecule collisions with $N>3$ requires heavy calculations of higher-order clusters. We leave this task for future work. 

The bound-state energies presented in Fig.~\ref{Fig:Spectra} are calculated by the stochastic variational method (SVM) \cite{SVMbook} where we set $r_0=0.003/\sqrt{m_{\rm F}|E_{\rm FB}|}$. In order to check universality of the results, we have performed additional runs with $r_0\sqrt{m_{\rm F}|E_{\rm FB}|}=0.006$ and 0.012; the relative error for the cluster energies presented in Fig.~\ref{Fig:Spectra} is at most 1\%. Our result for  $E_{\rm FBB}$ agrees with Ren and Aleiner \cite{RenAleiner2017} who have calculated and established universality of this quantity in the mass-balanced case, although in a narrower window of $a_{\rm BB}/a_{\rm FB}$.

Our calculations of the FFBB system are consistent with the qualitative guess that the $p$-wave molecule-molecule resonance should appear for $m_{\rm F}/m_{\rm B}\sim 1$. Indeed, for $m_{\rm F}/m_{\rm B}=2$ we find one tetramer state, which crosses the molecule-molecule threshold (corresponding to the molecule-molecule $p$-wave resonance) at $a_{\rm BB}/a_{\rm FB}=1.2 \times 10^{-5}$ [see Fig.~\ref{Fig:Spectra}(b)]. Although no FFBB tetramer state is found in the mass-balanced case, we anticipate a significant $p$-wave molecule-molecule attraction there, which is interesting for the realization of the topologically nontrivial (weakly coupled) $p_x+ip_y$ phase. We will now discuss scattering properties of the molecules.

The physical range of the molecule-molecule interaction is $a_{\rm FB}$ and, for small relative momenta $k\ll 1/a_{\rm FB}$, the scattering is characterized by the $p$-wave scattering surface $S$ and effective range $\xi$ defined through the effective range expansion of the scattering phase shift \cite{Randeria1989}
\begin{equation}\label{ERE}
-\pi\cot\delta(k)=\frac{4}{k^2S}-\ln(k^2\xi^2)+O(k^2a_{\rm FB}^2).
\end{equation}
In order to find these parameters we calculate the energy of the FFBB system in an isotropic harmonic potential of frequency $\omega$ and use the relation between the phase shift $\delta$ and the spectrum for two trapped particles (in our case FB molecules) derived by Kanjilal and Blume \cite{Kanjilal2006}
\begin{equation}\label{Relation}
-\pi\cot\delta(k)=\frac{1}{k^2l_0^2}-\ln\frac{k^2l_0^2}{2}+\psi\left(1-\frac{k^2l_0^2}{2}\right)+O(k^2a_{\rm FB}^2),
\end{equation}
where $l_0=1/\sqrt{(m_{\rm F}+m_{\rm B})\omega}$ is the oscillator length and $\psi$ is the digamma function. Eliminating $\delta$ from Eqs.~(\ref{ERE}) and (\ref{Relation}) and introducing $F(x,y)=x\psi(1-x/2y)+y+x\ln(2y)$ we arrive at the equation
\begin{equation}\label{EREFin}
F(k^2a_{\rm FB}^2,a_{\rm FB}^2/l_0^2)=\frac{4a_{\rm FB}^2}{S}-k^2a_{\rm FB}^2\ln\frac{\xi^2}{a_{\rm FB}^2}+O(k^4a_{\rm FB}^4).
\end{equation}
We emphasize that Eq.~(\ref{EREFin}) does not solve the four-body problem. It just relates the unknown scattering parameters $S$ and $\xi$ to the radial spectrum in a sufficiently shallow trap, which we calculate by the four-body SVM.

\begin{figure*}[hptb]
\begin{center}
\includegraphics[width=2\columnwidth]{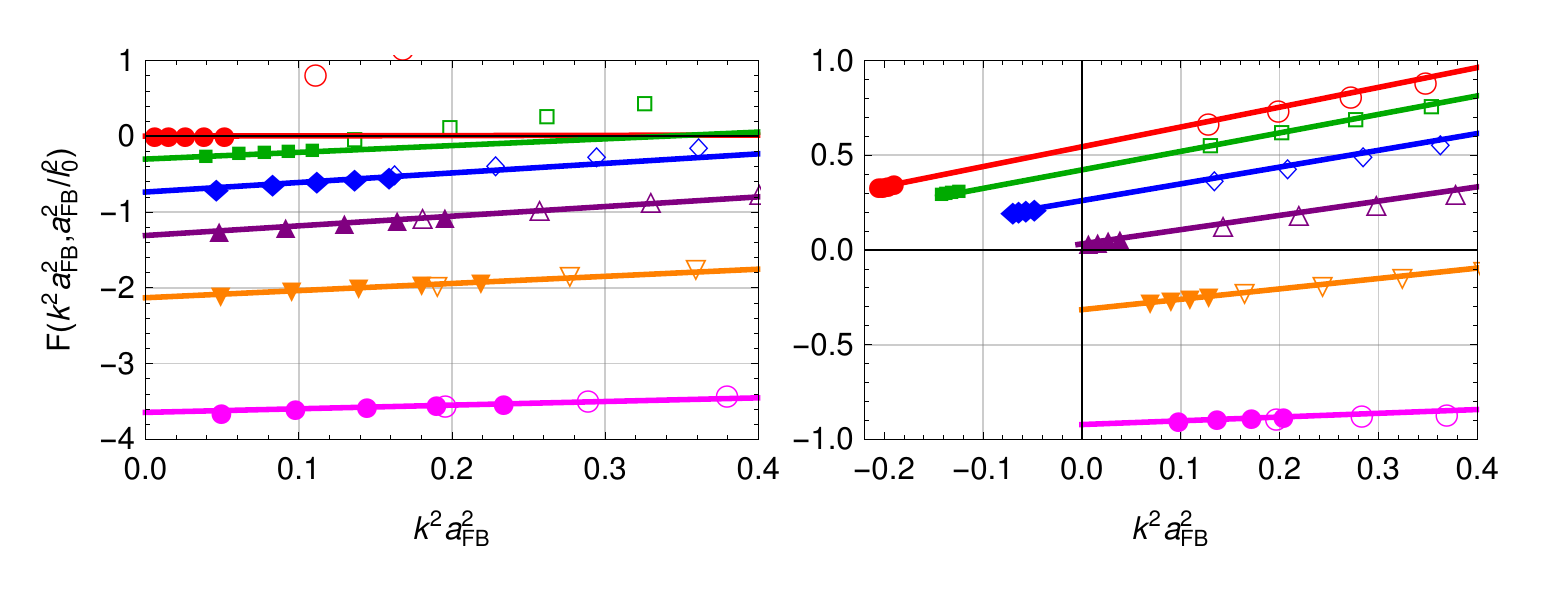}
\caption{$F(k^2a_{\rm FB}^2,a_{\rm FB}^2/l_0^2)$ vs $k^2a_{\rm FB}^2$ calculated for $a_{\rm BB}/a_{\rm FB}=10^{-8}$ (red), $10^{-7}$ (green), $10^{-6}$ (blue),  $10^{-5}$ (purple), $10^{-4}$ (orange), and $10^{-3}$ (pink) for $m_{\rm F}/m_{\rm B}=1$ (left) and $m_{\rm F}/m_{\rm B}=2$ (right). The full and hollow symbols correspond, respectively, to the ground and first radially excited four-body states in a trap with various $l_0$ (the leftmost data points stand for the largest $l_0$). Straight lines are linear fits to the ground-state data. Their left ends correspond to the free-space tetramer energies ($l_0=\infty$) shown in Fig.~\ref{Fig:Spectra}.}
\vspace{-0.5cm}
\label{Fig:Fit}
\end{center}
\end{figure*}

More specifically, we calculate the ground-state energy $E_{\rm FB}$ and the four-body spectrum $E_{\rm FFBB}$ as a function of $l_0$ for fixed $m_{\rm F}/m_{\rm B}$, $a_{\rm FB}$, and  $a_{\rm BB}$. The squared scattering momentum $k^2=(m_{\rm F}+m_{\rm B})(E_{\rm FFBB}-2E_{\rm FB})$ is plugged into $F(k^2a_{\rm FB}^2,a_{\rm FB}^2/l_0^2)$, which is then plotted versus $k^2a_{\rm FB}^2$ in Fig.~\ref{Fig:Fit}. The full and hollow symbols correspond, respectively, to the ground and first radially excited state of the relative molecule-molecule motion. The ground-state data are very well fit by linear functions (solid lines), from which we extract $S$ and $\xi$. For all data sets $0.5<\xi/a_{\rm FB}<1$. By contrast, $S$ can become large signaling a strong $p$-wave interaction, attractive for $S<0$. We see that the excited-state data (hollow symbols) are also consistent with the linear fits, although for $m_{\rm F}=m_{\rm B}$ and small $a_{\rm BB}/a_{\rm FB}$ we observe a deviation, which we attribute to the proximity of the FB-FB and F-FBB scattering thresholds (cf. Fig.~\ref{Fig:Spectra}).   

\begin{figure}
\begin{center}
\includegraphics[width=1.\columnwidth]{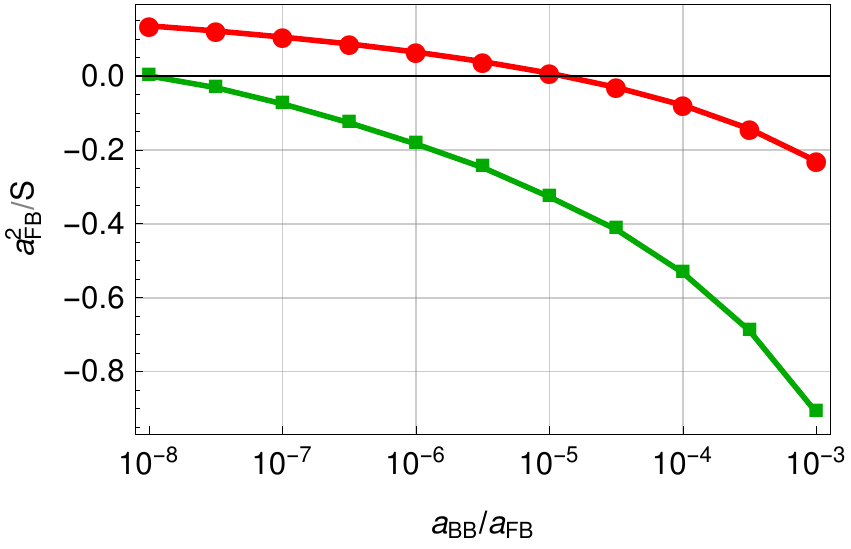}
\caption{\label{Fig:S}
$a_{\rm FB}^2/S$ vs $a_{\rm BB}/a_{\rm FB}$ for $m_{\rm F}/m_{\rm B}=1$ (green squares) and $m_{\rm F}/m_{\rm B}=2$ (red circles). }
\end{center}
\end{figure}

Figure~\ref{Fig:S} shows $a_{\rm FB}^2/S$ as a function of $a_{\rm BB}/a_{\rm FB}$ for the two mass ratios. The result is obtained for two molecules and in order to avoid cluster formation in three-molecule collisions, we need $a_{\rm BB}/a_{\rm FB}$ to be larger than $\approx 10^{-5}$ for both mass ratios (see Fig.~\ref{Fig:Spectra}). Nevertheless, even with this constraint we can access the resonance for $m_{\rm F}/m_{\rm B}=2$ (the resonance position is consistent with the tetramer threshold discussed earlier) and reach $S\approx -3a_{\rm FB}^2$ in the mass-balanced case, which is practically large given that $a_{\rm FB}$ is a controllable quantity much larger than the atomic interaction range. To give an idea for the value of the $p$-wave superfluid gap in a gas of such molecules, we cite the zero-temperature BCS result $\Delta\approx k_{\rm F}/[\xi(m_{\rm F}+m_{\rm B})]\exp(2/k_{\rm F}^2S)$ \cite{Randeria1989}, where $k_{\rm F}$ is the Fermi momentum. There are, of course, other observables  influenced by such a strong $p$-wave attraction: thermalization rate, mean-field interaction shift, virial coefficients, collective-mode frequencies, etc. Some of them are universally related through two $p$-wave contacts, corresponding to $S$ and $\xi$ \cite{Zhang2017contacts,Peng2016}. 

Favorable conditions for our proposal are provided by the isotopic $^{40}$K-$^{39}$K mixture. Confining it in one direction to zero-point motion results in a kinematically two-dimensional system characterized by the effective two-dimensional scattering lengths $a_{\rm FB}$ and $a_{\rm BB}$, related to the three-dimensional ones $a_{\rm FB}^{({\rm 3D})}$ and $a_{\rm BB}^{({\rm 3D})}$ by \cite{Petrov2001}
\begin{equation}\label{3Dto2D}
a=2e^{-\gamma}\sqrt{\pi/0.9}\;l_{\perp}e^{-\sqrt{\pi/2}l_{\perp}/a^{({\rm 3D})}},
\end{equation}
where $l_\perp$ is the confinement oscillator length. Near the $^{39}$K-$^{39}$K Feshbach resonance at 26G \cite{DErrico2007,Lysebo2010} one can tune $a_{\rm BB}^{({\rm 3D})}$ in a wide range keeping $a_{\rm FB}^{({\rm 3D})}\approx -20$nm \cite{SimoniPrivate,Cui}. Then, taking $a_{\rm BB}^{({\rm 3D})}=7$nm and $l_{\perp}=40$nm, which corresponds to the confinement frequency $\approx 2\pi\times 160$kHz achievable in current experiments \cite{LeticiaPrivate}, we obtain $a_{\rm FB}=1\mu$m and $a_{\rm BB}/a_{\rm FB}=7\times 10^{-5}$. Note that the FB molecules are confinement induced \cite{Petrov2001}, they require no FB Feshbach resonance and can be formed in a quasi-2D atomic mixture by three-body recombination \cite{Ngampuertikorn}. Their large size $a_{\rm FB} = 25 l_{\perp}$ ensures the applicability of our two-dimensional model.

Even though $a_{\rm BB}/a_{\rm FB}$ is above the threshold for the FBB trimer formation (see Fig.~\ref{Fig:Spectra}), collisions of two quasi-two-dimensional FB molecules can lead to the relaxation to deep molecular states of three-dimensional character. The process requires three atoms to approach one another to distances comparable to the van der Waals range (a few nanometers) where the trapping potential can be neglected and the reaction proceeds as the usual three-body recombination in three dimensions. Here the FBB relaxation channel is dominant since the FFB one is suppressed due to the Pauli exclusion. The corresponding three-body recombination rate constant in three dimensions is $K_3\sim (a_{\rm FB}^{(\rm 3D)})^4/m$,  where $m=m_{\rm F}\approx m_{\rm B}$ and we have just picked the largest of $a_{\rm FB}^{({\rm 3D})}$ and $a_{\rm BB}^{({\rm 3D})}$ \cite{RemarkDim}.

The FBB relaxation rate in molecule-molecule collisions for a given many-body state of the molecular gas is obtained by multiplying $K_3$ by the local FBB density correlator $\sim P_{\rm mm}G_{\rm FBB}$. Here, $P_{\rm mm}$ is the probability of finding a molecule (or, equivalently, a boson) in the immediate vicinity of another molecule, i.e., within a distance $\sim a_{\rm FB}$. This probability depends on the density of molecules, their typical momenta, and scattering properties ($S$ and $\xi$), and gets enhanced close to the resonance \cite{Levinsen2008}. In any case, for closely packed molecules ($k_{\rm F}a_{\rm FB}\sim 1$) $P_{\rm mm}$ saturates to one. 

The quantity $G_{\rm FBB}$ is the local FBB density correlator for a molecule and a boson confined axially to $l_\perp$ and longitudinally to a surface $\sim a_{\rm FB}^2$. Assuming the noninteracting FBB wave function we have $G_{\rm FBB}\sim {l_\perp}^{-2} a_{\rm FB}^{-4}$ and, accordingly, the relaxation rate $\nu = K_3 G_{\rm FBB}\sim [a_{\rm FB}^{(\rm 3D)}/l_\perp]^4 |E_{\rm FB}|(l_\perp/a_{\rm FB})^{2}$, which is much smaller than $|E_{\rm FB}|$. The main suppression comes from the factor $(l_\perp/a_{\rm FB})^{2} \ll 1$, relative to which $(a_{\rm FB}^{(\rm 3D)}/l_\perp)^4$ is only logarithmically small [see Eq.~(\ref{3Dto2D})]. One can show \cite{SM} that switching from the noninteracting to interacting FBB wave function when calculating $G_{\rm FBB}$ modifies only the logarithmic prefactor and leaves the main suppression coefficient unchanged. We thus claim that the loss rate of the quasi-two-dimensional molecules is much smaller than $|E_{\rm FB}|$. By contrast, the relaxation rate of weakly bound three-dimensional FB molecules in a similar closely packed configuration would be comparable to their binding energy ((see Supplemental Material \cite{SM}). Remarkably, these two equally strongly correlated systems, but in different dimensions, are so unequal from the viewpoint of three-body correlations. This phenomenon has the same physical origin as the presence (in three dimensions) or absence (in two dimensions) of the Efimov and Thomas effects for bosons \cite{Bruch1979,Lim1980}. 

In conclusion, weakly bound composite FB molecules experience a $p$-wave attraction provided by the boson exchange and counterbalanced by the fermion centrifugal barrier and Bose-Bose repulsion. The elastic and inelastic molecule-molecule scattering properties are sensitive to the dimensionality, Fermi-Bose mass ratio, and strength of the Bose-Bose interaction relative to the Fermi-Bose one. The quasi-two-dimensional isotopic $^{40}$K-$^{39}$K mixture is a promising and accessible candidate for obtaining a strongly $p$-wave-attractive Fermi gas. Further investigation of this system (in particular, characterization of its $p$-wave paired phase) can be performed by treating molecules as elementary fermions, interacting with each other by a convenient model potential (say, Gaussian) with properly tuned $S$ and $\xi$.

\begin{acknowledgments}
We thank L. Tarruell for useful discussions and A. Simoni for sharing his results on potassium scattering lengths. This research was supported by the European Research Council (FR7/2007-2013 Grant Agreement No. 341197).  
\end{acknowledgments}

\clearpage

\renewcommand{\theequation}{S\arabic{equation}}
\renewcommand{\thefigure}{S\arabic{figure}}

\setcounter{equation}{0}
\setcounter{figure}{0}

\onecolumngrid
\center{
\centerline{\underline{\bf SUPPLEMENTAL MATERIAL}}}
\vspace{5mm}
\twocolumngrid

\section{FBB correlator}

In the main text we argue that when we confine one fermion and two bosons to a quasi-two-dimensional slab $l_{\perp}\times a_{\rm FB}\times a_{\rm FB}$ and use the noninteracting three-body wave function to estimate the three-body correlator, the relaxation rate scales as 
\begin{equation}\label{Rate}
\nu \sim \frac{K_3} {{l_\perp}^{2} a_{\rm FB}^{4}}\sim \left[\frac{a_{\rm FB}^{(\rm 3D)}}{l_\perp}\right]^4 |E_{\rm FB}|\left(\frac{l_\perp}{a_{\rm FB}}\right)^{2}.
\end{equation}
However, the attractive FB interactions increase and the repulsive BB interaction decreases the local three-body correlator. We assume that the effect of these interactions at distances smaller than $l_\perp$ (where the kinematics is essentially three-dimensional) has already been captured by the dependence of $K_3$ on the three-dimensional scattering lengths. Thus, we are mostly interested in the behavior of the three-body wave function at distances larger than $l_\perp$ where the problem is two dimensional. Here we will use the purely two-dimensional model and treat interactions in the zero-range approximation.

The technical derivation in the rest of this section leads to a rather obvious result, which we state right away. At distances much larger than $a_{\rm BB}$ and much smaller than $a_{\rm FB}$ interactions are weak and the three-body wave function can be written in the Jastrow product form
\begin{equation}\label{Jastrow}
\Psi \propto \ln(r_{\rm FB1}/a_{\rm FB})\ln(r_{\rm FB2}/a_{\rm FB})\ln(r_{BB}/a_{\rm BB}),
\end{equation}
where $r_{{\rm FB}i}$ is the (longitudinal) distance from the fermion to the $i$-th boson and $r_{\rm BB}$ is the distance between the bosons. Equation~(\ref{Jastrow}) immediately gives the result claimed in the paper. Namely, interactions modify the rate (\ref{Rate}) by powers of logarithmic terms $\ln(l_\perp/a_{\rm BB})$ and $\ln(a_{\rm FB}/l_\perp)$.  

Let us denote the two bosons by indices 1 and 2, the fermion by 3, and introduce the mass-scaled two-dimensional Jacobi coordinates
\begin{align}\label{x}
&{\bf x}=\sqrt{\frac{4m_{\rm F}m_{\rm B}}{m_{\rm F}+2m_{\rm B}}}\left({\bf r}_3-\frac{{\bf r}_1+{\bf r}_2}{2}\right),\\
&{\bf y}=\sqrt{m_{\rm B}}({\bf r}_1-{\bf r}_2).\label{y}
\end{align}
For zero total angular momentum the three-body wave functions depends only on the hyperradius $\rho=\sqrt{x^2+y^2}$, hyperangle $\theta=2\arctan (|y|/|x|)$, and angle $\phi$ between vectors ${\bf x}$ and ${\bf y}$. In these coordinates the Schr\"odinger equation reads
\begin{widetext}
\begin{equation}\label{SchrHyper}
\left[-\frac{\partial^2}{\partial \rho^2}-\frac{3}{\rho}\frac{\partial}{\partial \rho}+\frac{4}{\rho^2}\left(-\frac{\partial^2}{\partial\theta^2}-\cot\theta \frac{\partial}{\partial\theta}-\frac{1}{\sin^2\theta}\frac{\partial^2}{\partial \phi^2}\right)-E\right]\Psi=0.
\end{equation}
The interaction between bosons corresponds to $\theta=0$ and sets the boundary condition at vanishing $y$ (or $\theta$)
\begin{equation}\label{BPBB}
\Psi \propto \ln\frac{|y|}{\sqrt{m_{\rm B}}a_{\rm BB}}\approx \ln\frac{\rho\theta}{2\sqrt{m_{\rm B}}a_{\rm BB}}.
\end{equation}
The interactions between bosons and fermions correspond to the points $\{\theta=\theta_0,\phi=0\}$ and $\{\theta=\theta_0,\phi=\pi\}$, where $\theta_0=2\arctan\sqrt{1+2m_{\rm B}/m_{\rm F}}$. The boundary conditions for the wave function at these points are identical up to the interchange $\phi\leftrightarrow \pi-\phi$. For $\phi=0$ we have
\begin{equation}\label{BPFB}
\Psi \propto \ln\frac{|{\bf r}_1-{\bf r}_3|}{a_{\rm FB}}\approx \ln\frac{\rho(\theta-\theta_0)}{2\sqrt{2\mu}a_{\rm FB}},
\end{equation}
where $\mu=m_{\rm F}m_{\rm B}/(m_{\rm F}+m_{\rm B})$.

In the adiabatic hyperspherical formalism we fix $\rho$ and diagonalize the angular kinetic energy operator in Eq.~(\ref{SchrHyper}) 
\begin{equation}\label{AngularSchr}
\left[-\frac{\partial^2}{\partial\theta^2}-\cot\theta \frac{\partial}{\partial\theta}-\frac{1}{\sin^2\theta}\frac{\partial^2}{\partial \phi^2}-\lambda(\rho)\right]\chi_{\lambda,\rho}(\theta,\phi)=0,
\end{equation}
with $\chi_{\lambda,\rho}$ satisfying Eqs.~(\ref{BPBB}) and (\ref{BPFB}). This problem is solvable analytically. For $\phi=0$ we have
\begin{equation}\label{chi}
\chi_{\lambda,\rho}(\theta,\phi)=C_{\rm BB}\chi^{(0)}(\theta)+C_{\rm FB}\chi^{(0)}(\theta')+C_{\rm FB}\chi^{(0)}(\theta''),
\end{equation}
where $\theta'(\theta)$ is the polar angle in the coordinate system with $z$-axis pointing at $\{\theta=\theta_0,\phi=0\}$, $\theta''(\theta)$ is the polar angle in the coordinate system with $z$-axis pointing at $\{\theta=\theta_0,\phi=\pi\}$, and 
\begin{equation}\label{chi0}
\chi^{(0)}(\theta)= \frac{\pi}{2}\tan\frac{\pi\sqrt{1+4\lambda}}{2}{\rm P}_{\sqrt{1/4+\lambda}-1/2}(\cos\theta)+{\rm Q}_{\sqrt{1/4+\lambda}-1/2}(\cos\theta)
\end{equation}
is the linear combination of the Legendre $P$ and $Q$ functions, that is smooth everywhere except $\theta=0$ where it behaves as
\begin{equation}\label{chi0smalltheta}
\chi^{(0)}(\theta)=-\ln(\theta/2)-\gamma-[\psi(1/2+\sqrt{1/4+\lambda})+\psi(1/2-\sqrt{1/4+\lambda})]/2 + O(\theta^2\ln \theta).
\end{equation}
Here $\psi$ is the digamma function. Substituting Eq.~(\ref{chi}) into Eqs.~(\ref{BPBB}) and (\ref{BPFB}) and using Eq.~(\ref{chi0smalltheta}) we obtain two linear homogeneous equations for the coefficients $C_{\rm BB}$ and $C_{\rm FB}$
\begin{align}\label{EqsForC1}
&\{\ln(\rho/\sqrt{m_{\rm B}}a_{\rm BB})-\gamma-[\psi(1/2+\sqrt{1/4+\lambda})+\psi(1/2-\sqrt{1/4+\lambda})]/2\}C_{\rm BB}+2\chi^{(0)}(\theta_0)C_{\rm FB}=0,\\
&\{\ln(\rho/\sqrt{2\mu}a_{\rm FB})-\gamma-[\psi(1/2+\sqrt{1/4+\lambda})+\psi(1/2-\sqrt{1/4+\lambda})]/2+\chi^{(0)}(2\pi-2\theta_0)\}C_{\rm FB}+\chi^{(0)}(\theta_0)C_{\rm BB}=0.\label{EqsForC2}
\end{align}

We are interested in the case $a_{\rm BB}\ll a_{\rm FB}$. Then, in the region 
\begin{equation}\label{region}
\sqrt{m_{\rm B}}a_{\rm BB}\ll \rho \ll \sqrt{\mu}a_{\rm FB}
\end{equation}
the logarithms in Eqs.~(\ref{EqsForC1}) and (\ref{EqsForC2}) are large. This physically means that the BB and FB interactions are, respectively, weakly repulsive and weakly attractive. We thus expect $\lambda(\rho)$ to be small in the region (\ref{region}). Indeed, expanding $\psi$ and $\chi^{(0)}$ at small $\lambda$, the consistency condition for the system (\ref{EqsForC1}-\ref{EqsForC2}) reads
\begin{equation}\label{lambdaSol}
\lambda(\rho)\approx 1/\ln(\rho/\sqrt{2\mu}a_{\rm FB})+1/2\ln(\rho/\sqrt{m_{\rm B}}a_{\rm BB}),
\end{equation}
where neglected are higher powers of $1/|\ln(\rho/\sqrt{2\mu}a_{\rm FB})| \ll 1$ and $1/\ln(\rho/\sqrt{m_{\rm B}}a_{\rm BB})\ll 1$.

We can now use (\ref{lambdaSol}) as the potential energy surface to find the hyperradial part of the three-body wave function, which satisfies
\begin{equation}\label{RadialSchr} 
\left[-\frac{\partial^2}{\partial \rho^2}-\frac{3}{\rho}\frac{\partial}{\partial \rho}+\frac{4\lambda(\rho)}{\rho^2}-E\right]\Psi(\rho)=0.
\end{equation}
We are interested in energies $E\sim 1/\mu a_{\rm FB}^2$ and, therefore, can neglect $E$ in Eq.~(\ref{RadialSchr}) for $\rho$ belonging to the region (\ref{region}). Note that although the perturbation $\lambda(\rho)$ is weak, it acts in a wide region of $\ln\rho$. One can see this more clearly by introducing the new variable $z=\ln\rho$ and substituting $\Psi(\rho)=e^{-z}W(z)$ into Eq.~(\ref{RadialSchr}) which then transforms into
\begin{equation}\label{RadialSchrZ} 
\left[-\frac{\partial^2}{\partial z^2}+\frac{4}{z-\ln(\sqrt{2\mu}a_{\rm FB})}+\frac{2}{z-\ln(\sqrt{m_{\rm B}}a_{\rm BB})}+1\right]W(z)=0
\end{equation}
\end{widetext}
and which one can solve in the region (\ref{region}) analytically by using the semiclassical WKB approximation.

Alternatively, we can just note that the ansatz
\begin{equation}\label{PsiSol}
\Psi(\rho)\propto [\ln(\sqrt{2\mu}a_{\rm FB}/\rho)]^\alpha[\ln(\rho/\sqrt{m_{\rm B}}a_{\rm BB})]^\beta
\end{equation}
with $\alpha=2$ and $\beta=1$ solves Eq.~(\ref{RadialSchr}) to the leading order in $1/\ln$ [i.e., $1/\ln(\rho/\sqrt{2\mu}a_{\rm FB})$ and $1/\ln(\rho/\sqrt{m_{\rm B}}a_{\rm BB})$]. 

Equation~(\ref{PsiSol}) is the main result of this derivation. One can easily check that it is nothing else than the hyperradial part of the Jastrow product (\ref{Jastrow}) (to the leading order in $1/\ln$).

In the paper we are interested in the case $m_{\rm F}\sim m_{\rm B}$. For completeness we mention that for $m_{\rm F}\ll m_{\rm B}$ (two heavy bosons interacting with a light fermion) the right inequality of (\ref{region}) can be rewritten as the constraint on the distance between the heavy bosons $|{\bf r}_1-{\bf r}_2|\ll \sqrt{m_{\rm F}/m_{\rm B}}a_{\rm FB}$. At distances $\sqrt{m_{\rm F}/m_{\rm B}}a_{\rm FB}\ll |{\bf r}_1-{\bf r}_2|\ll a_{\rm FB}$ the dependence of $\lambda$ on $\rho$ becomes linear, so that the effective potential $4\lambda(\rho)/\rho^2$ is hydrogen-like (cf. Ref.~[62] in the main text).

Finally, let us comment on the three-dimensional case. Consider a gas of weakly-bound three-dimensional FB molecules of size $a_{\rm FB}^{(3D)}$ (much larger than the van der Waals range $R_{\rm vdW}$) and density $\sim [a_{\rm FB}^{(3D)}]^{-3}$. As in the two-dimensional case, the relaxation rate of such molecules is on the order of the relaxation rate in the three-body FBB system confined to the volume $[a_{\rm FB}^{(3D)}]^{3}$. In the case $a_{\rm BB}\ll a_{\rm FB}$ the three-body wave function at hyperradii $\sqrt{m_{\rm B}}a_{\rm BB}^{(3D)}\ll \rho \ll \sqrt{\mu}a_{\rm FB}^{(3D)}$ is proportional to $\rho^{-2}$ (see Ref.~[66] in the main text; we neglect the Efimov log-periodic oscillations here) to be compared to the scaling $\rho^0$ in the noninteracting case. This means that interactions enhance the probability of finding three atoms in the recombination region $\rho \sim R_{\rm vdW}$ by $\sim [a_{\rm FB}^{(3D)}/R_{\rm vdW}]^4$ leading to relaxation rates comparable to the molecule binding energy. Whether the lifetime of this system can be increased by introducing a BB repulsion is an open question (see Ref.~[69] in the main text).  


\begin{thebibliography}{99}
\bibitem{remark1} 
It is obviously not essential whether Li is bosonic or fermionic here.
\end{thebibliography}
\end{document}